\documentclass[conference]{IEEEtran}
\IEEEoverridecommandlockouts
\usepackage{cite}
\usepackage{amsmath,amssymb,amsfonts}
\usepackage{algorithmic}
\usepackage{graphicx}
\usepackage{textcomp}
\usepackage{multirow}
\usepackage{circledsteps}
\pgfkeys{/csteps/fill color=aliceblue}
\usepackage{listings}
\usepackage{xcolor}
\usepackage{colortbl}
\usepackage[export]{adjustbox}
\usepackage[utf8]{inputenc}
\usepackage[english=american]{csquotes}
\usepackage{tabularx}
\usepackage{subcaption}
\usepackage{tcolorbox}
\definecolor{aliceblue}{rgb}{0.94, 0.97, 1.0}
\usepackage{comment}
\usepackage{url}
\usepackage{balance}
\usepackage[ruled,vlined]{algorithm2e}
\usepackage{graphicx}
\usepackage{footnote}
\usepackage[perpage]{footmisc}

\usepackage{cancel}
\usepackage{soul}

\usepackage{tcolorbox}

\newtcolorbox{conclusionbox}{
  colback=white,
  boxsep=1pt,
  arc=5pt,
  boxrule=0.5pt,
  width=\linewidth,
}

\newcommand{\name}[1]{\textit{RLExplorer}}

\newcommand{\gray}{\cellcolor{lightgray}}

\def\BibTeX{{\rm B\kern-.05em{\sc i\kern-.025em b}\kern-.08em
    T\kern-.1667em\lower.7ex\hbox{E}\kern-.125emX}}
\begin{document}

\title{Toward Debugging Deep Reinforcement Learning Programs with \name{}
}



\author{
\IEEEauthorblockN{Rached Bouchoucha*}
\textit{rached.bouchoucha@polymtl.ca}
\IEEEauthorblockA{
\textit{Mila, Polytechnique Montr\'{e}al, Canada}}
\and
\IEEEauthorblockN{Ahmed Haj Yahmed*}
\textit{ahmed.haj-yahmed@polymtl.ca}
\IEEEauthorblockA{
\textit{Mila, Polytechnique Montr\'{e}al, Canada}}
\and
\IEEEauthorblockN{Darshan Patil}
\textit{darshan.patil@mila.quebec}
\IEEEauthorblockA{
\textit{Mila, Universit\'{e} de Montr\'{e}al, Canada}}
\and
\IEEEauthorblockN{Janarthanan Rajendran}
\textit{rjana@umich.edu}
\IEEEauthorblockA{
\textit{Mila, Universit\'{e} de Montr\'{e}al, Canada}}
\and
\IEEEauthorblockN{Amin Nikanjam}
\textit{amin.nikanjam@polymtl.ca}
\IEEEauthorblockA{
\textit{Polytechnique Montr\'{e}al, Canada}}
\and
\IEEEauthorblockN{Sarath Chandar}
\textit{sarath.chandar@mila.quebec}
\IEEEauthorblockA{
\textit{Mila, Polytechnique Montr\'{e}al, Canada}}
\and
\IEEEauthorblockN{Foutse Khomh}
\textit{foutse.khomh@polymtl.ca}
\IEEEauthorblockA{
\textit{Polytechnique Montr\'{e}al, Canada}}
}


\maketitle
\def\thefootnote{*}\footnotetext{Both authors contributed equally to this research.}\def\thefootnote{\arabic{footnote}}

\begin{abstract}
Deep reinforcement learning (DRL) has shown success in diverse domains such as robotics, computer games, and recommendation systems. However, like any other software system, DRL-based software systems are susceptible to faults that pose unique challenges for debugging and diagnosing. These faults often result in unexpected behavior without explicit failures and error messages, making debugging difficult and time-consuming. Therefore, automating the monitoring and diagnosis of DRL systems is crucial to alleviate the burden on developers. In this paper, we propose \name{}, the first fault diagnosis approach for DRL-based software systems. \name{} automatically monitors training traces and runs diagnosis routines based on properties of the DRL learning dynamics to detect the occurrence of DRL-specific faults. It then logs the results of these diagnoses as warnings that cover theoretical concepts, recommended practices, and potential solutions to the identified faults. We conducted two sets of evaluations to assess \name{}. Our first evaluation of faulty DRL samples from Stack Overflow revealed that our approach can effectively diagnose real faults in 83\% of the cases. Our second evaluation of \name{} with 15 DRL experts/developers showed that (1) \name{} could identify 3.6 times more defects than manual debugging and (2) \name{} is easily integrated into DRL applications.

\end{abstract}

\begin{IEEEkeywords}
Deep Reinforcement Learning, DRL Bugs, Fault Diagnosis, Software tools, Software Engineering.  
\end{IEEEkeywords}



\section{Introduction}
Reinforcement learning (RL) is a branch of Machine Learning (ML) that focuses on autonomous learning and decision-making relying on interaction with an environment \cite{sutton2018reinforcement}. RL is based on trial and error. An agent interacts with its environment and learns to improve its behavior to achieve an objective expressed through scalar rewards \cite{sutton2018reinforcement, li2017deep, arulkumaran2017deep}. Deep reinforcement learning (DRL), harnessing Deep Learning (DL) in RL, has demonstrated promising capabilities in a variety of disciplines in recent years, such as robotics \cite{8675643}, computer games \cite{shao2019survey}, recommendation systems \cite{chen2021survey}, and computer vision \cite{panzer2022deep}. These accomplishments were made possible by DL, which enabled RL to scale to previously unreachable domains \cite{arulkumaran2017deep}. 
 
DRL systems, like other software systems, contain their unique faults \cite{nikanjam2022faults, zolfagharian2023search}, presenting developers/researchers with new challenges in debugging and diagnosing these systems. Developers are continuously releasing industrial-scale frameworks and libraries such as Stable Baselines \cite{stable_baseline}, Keras-RL \cite{plappert2016kerasrl}, TensorForce \cite{tensorforce}, and RL-Hive \cite{rl_hive} to aid practitioners in the design of reliable DRL systems. However, Debugging DRL systems is particularly challenging \cite{jones_debugging_nodate, deshpande2020interactive, rajan2019mdp} because, unlike conventional software systems, the decision logic of a DRL system is not explicitly encoded but rather derived from the interaction between the DRL agent and its environment \cite{ghosh2021generalization}. In addition, Deep Neural Networks (DNNs), the backbone of any DL/DRL system, are ``black box" entities that cannot be debugged using conventional methods such as breakpoints. To make matters worse, when faults are introduced, DRL systems often yield unexpected behavior without explicit error messages, or failures, making debugging tedious and time-consuming. Newcomers to RL, who frequently rely on high-level frameworks (like Stable Baselines \cite{stable_baseline}), have limited knowledge of debugging. RL experts, on the other hand, rely on their experience for debugging, often resorting to trial-and-error methods guided by intuition.

Currently, existing DRL frameworks offer limited aid for debugging faults during DRL system development. Tools like Tensorboard \cite{mane2015tensorboard}, Weights and Biases (Wandb) \cite{weights}, and tfdbg \cite{cai2017debug} can aid developers in monitoring DRL training. They are, however, incapable of analyzing the behavior of agents and providing meaningful insights. To address the aforementioned challenges, an approach that frees developers from manual monitoring and diagnosing DRL systems is paramount.

According to the DRL taxonomy of real faults proposed by Nikanjam et al., \cite{nikanjam2022faults}, faults in DRL systems are divided into three categories, and in this paper, we focus on RL-specific faults (e.g., missing exploration), and DNN-specific faults (e.g., vanishing gradient). The third category, generic programming faults, was excluded from our study due to its prior investigation in other related studies~\cite{humbatova2020taxonomy, islam2019comprehensive}. We propose \name{}, the first fault diagnosis approach for DRL systems. To design \name{}, an initial literature review was conducted to identify diagnostic routines targeting faults in DNNs. Our focus converged on the work of Ben Braiek and Khomh \cite{braiek2022testing}, which offers a comprehensive list of DNN-related diagnosis routines. Next, we selected and adapted these routines for implementation within DRL systems, thereby enriching the diagnostic toolbox available for DRL systems. Second, we leverage Nikanjam et al.’s taxonomy \cite{nikanjam2022faults} of DRL-specific training faults. Then, we identify symptoms and root causes that are generated by these DRL training faults from prior AI research \cite{xin2020exploration, jones_debugging_nodate, lockwood2022review} and introduce new diagnosis routines that detect their occurrences during training. Attached to the DRL application runtime, \name{} leverage dynamic analysis \cite{myers2011art} to automatically record data and run diagnosis routines in the form of checks that encapsulate properties of the learning dynamics. \name{}, then, displays the results of these checks as warning messages that incorporate theoretical concepts and explain best practices to address the root causes. 
For example, assume the Action Stagnation symptom, which occurs when the agent gets stuck in a local optimum \cite{hong2018diversity}. In that case,  \name{} would identify this symptom by monitoring the number of identical actions within a single episode or between two successive action sequences.

To evaluate \name{}, we assessed \name{} using 11 real faulty DRL samples collected from Stack Overflow (SO). Results show that \name{} was able to detect and diagnose faults in 83\% of cases. We further conduct a human study with 15 DRL practitioners to assess \name{}'s effectiveness and usability in helping developers diagnose faults. Participants using \name{} were able to diagnose 3.6 times more faults compared to manual debugging. Participants also reported high satisfaction with the debugger and a high likelihood of leveraging \name{} in their development.

In summary, our work makes the following contributions:
\begin{itemize}
    \item \textbf{Novelty}: We propose \name{}, the first fault diagnosis approach for DRL systems. \name{} automatically runs diagnosis routines based on learning dynamics traces to detect the occurrence of RL and DNN faults.

    \item \textbf{Utility}: \name{} automatically records data and runs checks that cover learning dynamics. It then displays the checks' results as warnings that combine theoretical concepts and explain best practices. \name{} supports on- and off-policy model-free RL algorithms.

    \item \textbf{Community involvement}: We collected and reproduced 11 faulty programs derived from SO. This bug collection acts as the ground truth for evaluating our approach and can be used in future research on DRL debugging and repair. We also provide our replication package \cite{blind_toward_2023} to promote future research and investigations.

    \item \textbf{Evaluation}: First, we assessed \name{} on 11 SO faulty samples. The result shows that \name{} can diagnose faults for 83\% of cases. Second, we conducted a study with 15 DRL experts to assess \name{}'s effectiveness and usability. Participants using \name{} found 3.6 times more faults than manual debugging and found \name{} easy to use in DRL applications.
\end{itemize}


\section{Background}
\label{sec:background}

\subsection{Deep Reinforcement Learning}
\label{sec:DRL}
Standard RL \cite{kaelbling1996reinforcement} uses the framework of Markov Decision Processes (MDP) \cite{white1993survey} to define the sequential decision-making problem. In RL an MDP can be represented as
a tuple $(\mathcal{S}, \mathcal{A}, \mathcal{T}, \mathcal{R}, \gamma)$ 
where $\mathcal{S}$ is the set of states,  
$\mathcal{A}$ is the set of possible actions,  
$\mathcal{T}(s_{t+1}| s_t, a_t)$ represents the transition dynamics that describe the probability distribution over states at time $t+1$ given a state-action pair at time $t$, $\mathcal{R}:\mathcal{S}\times\mathcal{A} \rightarrow \mathbb{R}$ is a reward function,
and $\gamma \in [0,1]$ is a discount factor where lower values denote a preference for more immediate rewards.\\ 
A policy $\pi(a|s)$ defines a distribution over actions given a state.
In general, RL agents try to learn policies that maximize the discounted return at each step:
\begin{equation}
    G_t = R_{t+1} + \gamma R_{t+1} + \gamma^2 R_{t+2} + \cdots = \sum_{k=0}^\infty \gamma^k R_{t+k+1}
\end{equation}
Where $G_t$ is the discounted return and $R_t$ is the immediate reward at timestep $t$.
The expected discounted return given action $a$ is taken at state $s$ and the agent is following policy $\pi$ which is known as a Q-function:
\begin{equation}
\label{eq:Qfunction}
    Q_\pi(s,a) = \mathbb{E}_\pi[G_t | S_t=s, A_t=a].
\end{equation}

Most RL algorithms use function approximators to learn the optimal policy, $\pi_*$ \cite{arulkumaran2017deep}, or $q_*$, the Q-function \cite{watkins1992q} of the optimal policy. Tabular function approximators make learning these functions in high-dimensional, continuous observation or action spaces difficult or impossible. 
DRL \cite{li2017deep}\cite{arulkumaran2017deep}, RL with DNN function approximators, has emerged as one of the most active areas of research in ML.
DNN function approximators allow the use of RL methods for complex state and action spaces, overcoming the limitations of traditional RL.
Training a DRL agent presents challenges such as ensuring effective exploration of the environment \cite{ladosz2022exploration} and stabilizing the training of the function approximators used by the agent \cite{li2017deep}\cite{packer2018assessing}.

\subsection{Differences in diagnosing DRL over other DL programs}
In addition to the issues with training NNs faced in standard DL settings\cite{wardat2022deepdiagnosis, wardat2021deeplocalize, cao2022deepfd}, DRL systems also face unique challenges.
While standard DL settings usually involve training a model on a static dataset, the data used to train the model in RL is generated as a function of a changing agent and a dynamic environment. This has several implications: 
(1) To even access all the relevant experience that an environment provides, the agent must be able to explore well; otherwise, the agent will likely fail as it will encounter states it has not seen before.
(2) As the agent learns, the experience it generates---and thus the data it is trained on---changes. For example, as an agent learns, it might be able to reach different parts of the environment which not only leads to new states the agent must learn but also potentially having to change its evaluation of states from earlier in the training process. 
(3) Unlike in other settings where the loss is expected to (generally) monotonically decrease, in RL, because of the non-stationary nature of the data, the trend of the loss does not necessarily correspond to the agent performance.
(4) Due to the complex, dynamic nature of the system, there is often more noise in the results of RL experiments. Usually, most RL experiments need to be run for several seeds before any conclusions can be drawn.
Finally, since the DRL system consists of multiple dynamic components that the developer must implement, detecting faults when they occur can be challenging.


\section{Approach}
\label{sec:approach}
In this section, we explain our DRL-fault diagnosis approach, \name{}. We describe the list of the DRL faults that \name{} monitors, along with their associated symptoms and root causes. Then, we detail our approach to detect these faults and explain the workflow of \name{}. To the best of our knowledge, \name{} is the first approach that thoroughly addresses the challenge of fault diagnosis specifically in DRL.

\subsection{Overview}

\begin{figure*}[h]
\centering
\includegraphics[width=.95\textwidth]{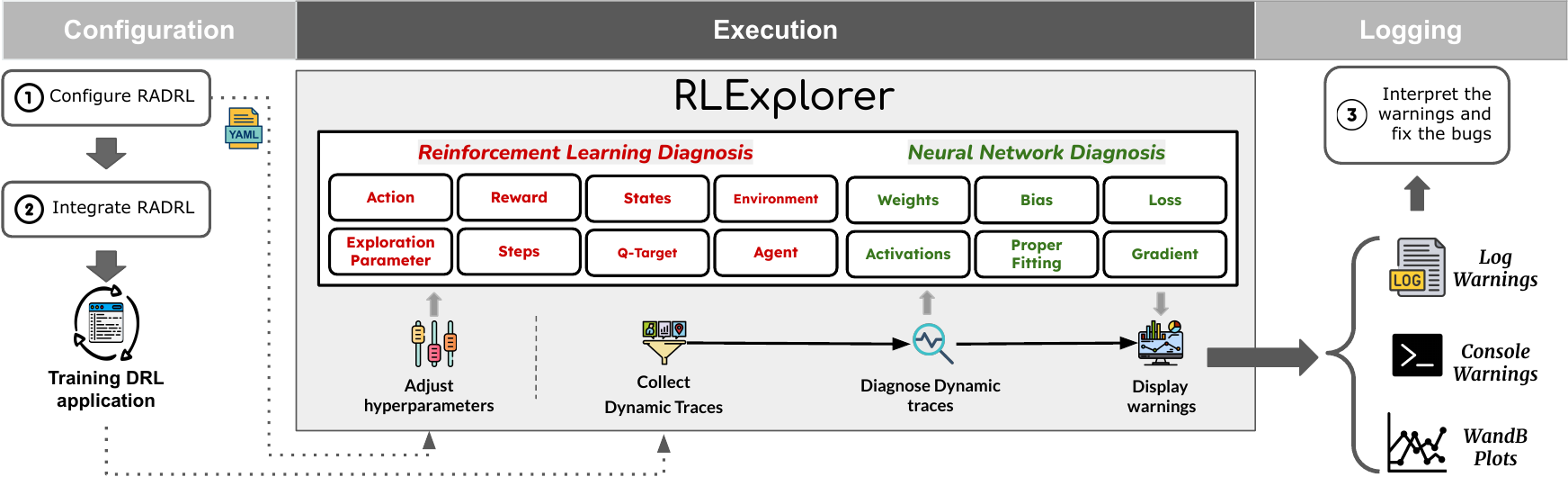}
\caption{Illustration the Three Steps of \name's Workflow: Configuration, Execution, and Fault Logging.}
\label{fig:framework}
\vspace{-1em}
\end{figure*}
Training DRL applications is known to be time-consuming and computationally expensive \cite{duan2016benchmarking}\cite{clemente2017efficient}. \name{}’s main objective is to identify symptoms of faults as soon as possible during training. This can help DRL developers save time/effort and find faults without waiting until the end of the training.
Figure \ref{fig:framework} illustrates an overview of \name{}. \name{} begins by collecting various dynamic traces (e.g., reward, weights, and actions) from the DRL application during the training. It then analyzes these traces for automatic and real-time identification of potential faults. Finally, if a symptom is detected, \name{} prompts warning messages that guide the user to address the root causes of the issue. \name{} has three phases: Configuration, Execution, and Logging. 

\paragraph{Configuration}

\name{} offers a set of diagnostics to check the behavior of key components in a DRL application. Users specify which components to diagnose by selecting relevant diagnostics and their execution frequency. Each individual diagnostic can be further customized to meet the training environment's specific characteristics. Besides, to enhance the \name{}'s usability, a default configuration is provided.

\paragraph{Execution}

\name{} monitors learning traces to perform the diagnosis. The \name{}'s dynamic trace collection phase employs two methods: (1) \name{} automatically detects standard DRL components’ traces (e.g., rewards and states from the environment) using the observer design pattern, and (2) for traces that cannot be detected automatically (e.g., exploration parameter), \name{} provides a predefined function that users can call to track these traces. Then, each chosen diagnostic periodically performs its verification routines looking for any symptom of a specific component.

\paragraph{Logging}

During training, \name{} operates as an event handler. When a symptom is detected by a diagnostic, \name{} is triggered to process the symptom and notify the user when a fault is detected. \name{} displays warning messages that provide explanations of the faulty behavior and best practices to address the root cause. Finally, \name{} provides real-time plots of advanced features (e.g., epistemic uncertainty \cite{da2020uncertainty}) that conventional monitoring tools like Tensorboard \cite{tensorboard} do not show. The event-driven nature of \name{} enables developers to resolve errors promptly rather than waiting until the end of the training process. This is crucial in DRL applications given that training is a resource-intensive and time-consuming task.
\subsection{Failure symptoms and root causes} 
\label{sec:symp}
To establish the diagnostic approach of \name{}, we first identify symptoms of DRL faults and their root causes from prior AI research \cite{nikanjam2022faults, xin2020exploration, jones_debugging_nodate, lockwood2022review}. Subsequently, we developed dynamic diagnosis strategies to identify these failure symptoms and suggest possible mitigation actions (Section \ref{sec:detfaisymp}). In the following subsections, we explain the DRL faults symptoms, their underlying causes, and the methods employed by \name{} to detect such symptoms. Notably, we classified DRL faults into two categories: RL-specific faults and NN-specific faults.\\

\subsubsection{The Reinforcement Learning Symptoms}
\label{sec:rlsymp}
Nikanjam et al. \cite{nikanjam2022faults}, proposed a taxonomy of DRL faults based on an analysis of SO posts and GitHub issues to identify commonly reported faults among DRL developers. While this taxonomy served as a starting point for our work, the primary contribution of \name{} is the creation of fault diagnosis routines. These routines are the result of an extensive analysis and adaptation process, aiming to address each fault category and detect its symptoms.  Each fault may manifest in the behavior of various RL components. In the following, we present the RL-specific symptoms and root causes for 8 RL components.

\noindent{\textbf{Actions Symptoms.}} \\
\underline{Abnormal state entropy:} Xin et al. \cite{xin2020exploration} define State Entropy (SEN) as the RL agent's uncertainty and randomness in selecting an action at each state. SEN tends to start high (during exploration) then decreases throughout learning and reaches its minimum when the learning process converges, providing the best-learned policy \cite{jones_debugging_nodate}. SEN's erroneous behavior symptoms might include prolonged stagnation, sharp drops, unexpected increases, or severe fluctuations \cite{jones_debugging_nodate}. \textit{Root causes:} Prolonged SEN stagnation can occur when the policy targets or gradient backpropagation are incorrectly computed \cite{jones_debugging_nodate}. Sharp drops and unexpected increases in SEN can occur when the agent collapses into a myopic policy (only considers the immediate reward \cite{wang2018deep}) and is not exploring anymore. Finally, severe fluctuations occur when the learning rate is too high \cite{jones_debugging_nodate}. \\
\underline{Action stagnation:} An agent repeating the same sequence of actions is a symptom of faulty behavior. Stagnation can occur inside an episode when the agent repeats the same action, or across successive episodes when the agent repeats the same sequence of actions. \textit{Root causes}: Action stagnation may be caused by performance plateau \cite{schaul2019ray} when the agent gets stuck in a local optimum \cite{hong2018diversity}, or by the Noisy TV problem \cite{burda2018exploration}. \\
\underline{High epistemic uncertainty (EU):} EU is a learner's uncertainty caused by a lack of knowledge during a learning task \cite{lockwood2022review, clements2019estimating}. It reveals the learner's (agent's) confidence in decision-making. Typically, the EU value is high in the early stages of training and is expected to decline as iterations and data grow \cite{lockwood2022review}. Any deviation from this expected behavior may indicate the presence of faulty behavior. \textit{Root causes:} High EU may be caused by a missing or suboptimal exploration or wrong computation of the agent’s gradient during backpropagation. 

\noindent{\textbf{Agent Symptoms.}} \\
\underline{Wrong agent update policy:} 
Leveraging DNNs as a function approximation in DRL makes the agent's learning unstable \cite{kobayashi2021t}. This is due to the DNN's nonlinearity. To that end, techniques have been introduced to stabilize the agent's learning in DRL \cite{kobayashi2021t}. In this vein, target networks were proposed by Mnih et al. \cite{mnih2015human} as a stable reference for the main network to reduce the learning variance. Thus, it is recommended to use a target model to stabilize the learning process. The wrong agent update symptom arises when the synchronization between the main and target networks is disrupted, causing an unstable learning process. \textit{Root causes}: This symptom is due to the mix-up between the main and target networks, not updating the target network at the right time, or not updating it at all.\\ 
\underline{High Kullback-Leibler (KL) divergence in predictions:} This symptom occurs when the KL divergence \cite{Kullback1987LETTERTT} of the main model predictions is high on a fixed batch of states over two consecutive model updates. The intuition is that the outputs of the model should remain consistent and not be changed drastically in few successive updates \cite{jones_debugging_nodate}. \textit{Root causes}: This symptom may arise if the agent learns from new observations with a divergent distribution \cite{jones_debugging_nodate}. 

\noindent{\textbf{Environment Symptoms.}}\\
\underline{Wrong environment conception:} When designing an environment it is crucial to ensure that the environment design conforms to the DRL standards \cite{brockman2016openai}. Symptoms of faulty environment design occur as numerical instabilities (NaN or infinity values) in the outputs of environmental functions (e.g. step and reset). \textit{Root causes}: The root causes are often a wrong design of the RL environment or missing the terminal state.

\noindent{\textbf{Exploration Parameter Symptoms.}}\\
\underline{Suboptimal exploration rate:} Effective learning strategies initiate with a high exploration rate to encourage the agent to explore different actions and states, gradually reducing the exploration rate to favor exploitation. Controlling this exploration-exploitation balance is often governed by a parameter, like $\epsilon$ in the epsilon-greedy policy \cite{rawson2021convergence}. Symptoms in this category manifest in a sharp decrease in this parameter. \textit{Root causes}: These symptoms can be caused by improper setup, update, or high variation in the exploration parameter.

\noindent{\textbf{Reward symptoms.}}\\
\underline{Erroneous accumulated reward behavior:} During early training, the DRL agent's behavior is mostly random (the agent is exploring), reflected by high and fluctuating standard deviations (std) in rewards. As training progresses, the agent's behavior becomes more stable, and the std of the accumulated reward should decrease. Any deviation from this expected trend could indicate an erroneous behavior. \textit{Root causes:} One likely cause of this symptom is a suboptimal exploration rate.

\noindent{\textbf{Step Symptoms.}} \\
\underline{Premature episode termination:} The number of steps executed by an agent while interacting with its environment directly impacts the agent's learning efficiency. Premature termination episodes can hinder this efficiency, especially during exploitation. \textit{Root causes:} A common cause is the low value set for the maximum number of steps per episode.

\noindent{\textbf{States Symptoms.}}\\
\underline{Repetitive states sequence:} Symptoms in this category include repetitive or stagnant state sequences that hinder the agent from achieving its goal. This can occur in two forms: First, repeated sequence of states within one episode, where the environment keeps returning to the same states. Second, repeated sequences of states across multiple episodes, where the agent repeatedly follows the same path. \textit{Root causes:} Repetitive states may indicate that the agent is stuck in a local optimum and requires further exploration \cite{hong2018diversity}.

\noindent{\textbf{Q-Target Symptoms.}} \\
\underline{Wrong calculation of the q\_targets:} This symptom concerns DRL applications using Q-value-based learning, where discrepancies arise between the q\_targets values, computed according to the original formula (\ref{eq:Qfunction}), and the one provided by the user. \textit{Root causes:} The main cause of this symptom is the wrong application of the q\_targets calculation formula. \\

 \subsubsection{The Neural Network (NN) Diagnoses}
The NN diagnostics involve checking various features of the NN such as weights, and biases during the model training. Several studies have proposed taxonomies of common faults in NNs \cite{humbatova2020taxonomy, chen2021empirical}, along with approaches for detecting and fixing these faults \cite{cao2022deepfd, wardat2022deepdiagnosis, wardat2021deeplocalize, braiek2022testing}. Building on the work of Ben Braiek and Khomh \cite{braiek2022testing}, who proposed DNN training faults, symptoms, and their diagnostic routines, we have selectively adapted these diagnostic routines to address DRL-specific needs. We adapted NN diagnoses that targeted DRL-compatible features such as \textit{activation}, \textit{bias}, \textit{gradient}, \textit{loss}, \textit{weight}, and\textit{ proper fitting} while omitting others like accuracy and labels unsuitable for DRL contexts. For more details on these adaptations, refer to \cite{braiek2022testing} and our replication package \cite{blind_toward_2023}.

\begin{table*}[htbp]
\centering
\caption{Description of \name{}'s Diagnosis Approach for Detecting Fault Symptoms in RL Components (\ref{sec:rlsymp})}
\label{tab:diagnosis}
\begin{tabular}{c l l l }
\hline
\textbf{Component} & \textbf{Input} & \textbf{Symptoms} & \textbf{Detection Method}\\

\hline
\rowcolor{lightgray}  & \textit{MM}: main model;& Wrong agent & If the current step has reached $TM_{per}$, we check if TM’s params are updated\\
\rowcolor{lightgray} Agent  & \textit{TM}: target model; &  update policy              &with MM (d1). If not, we check if the two model’s params are different (d2).  \\
\rowcolor{lightgray} (AGT) & \textit{AP}: actions probs;&                &We also check that APs are outputted by the correct model (MM) (d3).  \\
\rowcolor{lightgray}     & \textit{TM$_{\text{per}}$}: TM update &  High KL &Finally, in two successive updates, we check if the agent's output KL divergence\\
\rowcolor{lightgray}       & period;  & divergence   & \cite{Kullback1987LETTERTT} on a given input batch is smaller than $0.1$ (d4).  \\
\hline
Environment & \textit{env}: environment;& Wrong environment & We analyze the environment’s behavior against random actions. We check for numerical \\
(ENV) & & conception                               & instabilities (NaN or infinity values) (d1), and unnormalized rewards as recommended \\
      & &                                          & by \cite{jones_debugging_nodate} (d2). We also checks if random episodes can easily reach the episode's \\
      & &                                          & maximum reward threshold (too-easy problem) (d3). \\
\hline
\rowcolor{lightgray} States & \textit{env}: environment;& Repetitive states & We check for unnormalized observations outside [-10,10] range (recommended in \cite{jones_debugging_nodate}) \\
\rowcolor{lightgray} (STT) & \textit{Max$_{\text{r}}$}: max reward; & sequence & (d1). Next, in the last 20\% of episodes, we check for identical states within one \\
\rowcolor{lightgray}      & &                                          & episode (d2) or across successive episodes (d3). \\
\hline
Step & \textit{env}; \textit{Max$_{\text{r}}$}: max & Early episode & We check if episodes are prematurely ended due to Max$_{\text{stps}}^{\text{ep}}$ being reached whilst \\
(STP) & reward; \textit{Max$_{\text{stps}}^{\text{ep}}$}: max & termination & the agent fails to reach a reasonable average reward (e.g., 0.1 * Max$_{\text{r}}$) (d1). \\
      & steps per episode;&      & d1 only runs in the last 20\% episodes. \\
\hline
\rowcolor{lightgray} Exploration & \textit{EF}: exploration & Suboptimal & We check if the function of the Least Squares Solutions ($lstsq$) \cite{abdi2007method} of EF values is not \\
\rowcolor{lightgray} Parameter &factor; & exploration rate & strictly monotonous (d1). Next, we check if the second derivative of EF values ($f_{EF}''$) \\
\rowcolor{lightgray} (EXP)       & &                       & is too high ($> 0.22$), which indicates that the rate of change of EF values is fast (d2). \\
\hline
 & \textit{R$_{\text{std}}$}: reward std; & Erroneous & This check examines R$_{\text{std}}$ in a window of successive episodes. In the first 20\% episodes, \\
  & \textit{Max$_{\text{r}}$}: max reward; & accumulated reward & we check if the reward per episode is stagnating. For that, we verify if the RMSE \cite{chai2014root} \\
  Reward     &                   & behavior & between the Least Squares solutions ($lstsq$) of R$_{\text{std}}$ and the actual R$_{\text{std}}$ values is less  \\
   (RWD)   & &                         & than $0.1$ (d1). In the last 20\% episodes, we check if the slope of $lstsq$ is fluctuating \\
      & &                         & ($|lstsq| > 0.25$) (d2) and if the agent is trapped at a low reward value (d3). \\
\hline
\gray  & \gray  & \gray Abnormal state & \gray First, we check the State Entropy (SE). In the first 20\% episodes, we check if SE is \\
\gray   & \gray & \gray entropy (SE) & \gray increasing ($lstsq >0.1$)  (d1) or stagnating ($<10^{-3}$) (d2). In the last 20\% episodes, \\
\gray        & \gray                  & \gray & \gray we check the rate of change of the SE is fast (method in EXP.d2)(d3). During the whole \\
\gray       & \gray & \gray           & \gray training, we check if SE is fluctuating using the same method as RWD.d1 (d4). \\
\gray Actions     & \gray   \textit{AP}: actions probs; & Action stagnation  & We check if the number of identical actions within a single episode (d5) or between \\
\gray (ACN)    & \gray \textit{Max$_{\text{r}}$}: max reward; &   & two successive action sequences is ($>10$) (d6). \\
\gray       & \gray & \gray High epistemic  & \gray Monte Carlo dropout quantifies uncertainty by adding multiple dropout layers to the \\
\gray       & \gray & \gray uncertainty     & \gray network and generating different outputs for a given input. We check if the average  \\
\gray       & \gray & \gray                 &  \gray std of model outputs is too high ($>0.5$)(d7). \\
\hline
Q-Target & \textit{QTs}: q\_targets; & Wrong calculation & We check the equivalence of the q\_targets obtained using the equation \ref{eq:Qfunction} \\
(QTR)    & \textit{QTs$_{\text{pred}}$}:pred QT& of the q\_targets & with those provided by the user (d1). \\
\hline
     
\end{tabular}
\vspace{-1em}
\end{table*}

\subsection{Detecting Failure Symptoms}
\label{sec:detfaisymp}
In Table \ref{tab:diagnosis}, we report our approach for detecting the symptoms discussed in Section \ref{sec:symp}. Before running \name{}, users can customize the checks' configurations specified in Table \ref{tab:diagnosis}. We also established default configuration values through trial-and-error experimentation and validation with DRL domain experts. After various trials on different RL algorithms/environments, we believe that the given default configuration values demonstrated a satisfactory level of generalizability in different DRL problems.

After configuring and integrating \name{} in the DRL system, \name{} subscribes to the user-activated components and listens to their required dynamic traces (Inputs in Table \ref{tab:diagnosis}). Then, \name{} is periodically triggered to perform three types of diagnostics: early-stage learning diagnostics, late-stage learning diagnostics, and diagnostics performed throughout the training process. The early-stage diagnostics investigate symptoms that occur during exploration including ENV.d1, ENV.d2, ENV.d3, STT.d1, RWD.d1, ACN.d1, ACN.d2 (check Table \ref{tab:diagnosis}); 
while the late-stage diagnostics handle the symptoms that may occur during exploitation including STP.d1, STT.d2, STT.d3, RWD.d2, RWD.d3, ACN.d3. The remaining diagnostics (AGT.d1, AGT.d2, AGT.d3, AGT.d4, EXP.d1, EXP.d2, ACN.d4, ACN.d5, ACN.d6, ACN.d5, QTR.d1) are performed during the whole training. Noting that we did not cover NN diagnostics here due to space constraints, we suggest readers look at \cite{braiek2022testing} and our replication package \cite{blind_toward_2023} for more information on NN symptoms and methods for identifying them. Finally, if a symptom is detected, \name{} logs and notifies the user with a warning message that highlights the fault explanation. Additionally, \name{} provides real-time plots of the epistemic uncertainty, reward standard deviation, and KL divergence.

\subsection{Integration of custom fault detection diagnosis}
To enhance the flexibility of our proposed tool, we have focused on simplifying the process of integrating new fault diagnosis strategies, customized to the user's specific needs and requirements. This integration can be carried out programmatically in two simple steps: firstly, by implementing the user's fault diagnosis strategy, and secondly, by activating it. The tool manages the rest of the process autonomously.\\
The reason for this seamless integration lies in employing several strategic design patterns for the architectural design of our approach, such as Observer \cite{szallies1997using}, Factory \cite{raj1999factory}, Registry \cite{fowler2012patterns}, and Singleton \cite{gamma1995elements}, known for their ability to give applications greater flexibility and customizability. These patterns enable our tool to easily adapt to various fault diagnosis logics, increasing its overall usefulness and adaptability.

\section{Evaluation}
\label{sec:evaluation}
\label{sec:eval}
In this section, we aim to evaluate \name{} using (1) real faulty samples extracted from SO and (2) a study involving human participants. For the first evaluation, we collected and reproduced real faulty samples gathered from SO to assess \name{}'s effectiveness in fault diagnosis and reported the number of accurately diagnosed faulty cases. We also evaluated the runtime overhead that \name{} adds to the collected SO faulty samples. For the second evaluation, we conducted a human study through a coding task and survey to solicit DRL experts' feedback on \name{}.

\subsection{Faulty Samples evaluation}

\subsubsection{Method}
Using a mining strategy similar to prior works \cite{wardat2021deeplocalize, zhang2021autotrainer}, we collected and reproduced real DRL faulty samples from SO posts.
We started by searching for posts on SO with the tag ``reinforcement learning." We then retained posts with (i) accepted answers and (ii) code snippets in the post description. After removing duplicates and posts having the term ``install" (to exclude installation-related posts), we obtained a total of 426 posts. In the second phase, two authors manually examined the retrieved posts and filtered out those that included incomplete code, non-reproducible statements, or crashing errors. In the final phase, we reviewed each post's accepted answer, identified the faults, their symptoms, and root causes, and defined at least one expected fault diagnosis. The two authors performed cross-validation of the extracted posts to ensure agreement. Conflicts were handled and resolved through scheduled meetings.
In total, we collected 11 posts and 12 faults. The low number of posts is due to the lack of reproducible DRL faulty codes in SO. We also note that there is no available dataset of reproducible DRL buggy programs.

\subsubsection{Evaluation criteria}
We use the following criteria to determine if a fault is effectively detected and diagnosed by \name{}: A fault is deemed correctly diagnosed if \name{}'s detection result corresponds to the post's accepted answer. Additionally, we measured the time overhead introduced by \name{} when debugging DRL faulty samples using the following formula:
\begin{equation}
\text{Time overhead (\%)} = \frac{T_d - T_n}{T_n} \times 100
\end{equation}
where $T_d$ is the time taken to execute the program with the debugger enabled and $T_n$ is the time taken to execute the program without the debugger. Due to the stochasticity of DRL systems, we report the average time overhead of $5$ runs.

\subsubsection{Results}

\begin{table*}[htbp]
  \centering
  \caption{Results of \name{} on Stack Overflow Buggy Samples.}
  \label{tab:table1}
  \begin{tabular}{l l l c c c}
    \hline
    \textbf{Post \#} & \textbf{Ref} & \textbf{Description}  & \textbf{Runtime (s)}  & \textbf{Time}  & \textbf{Fault}  \\
                       &                      & \textbf{} &  & \textbf{Overhead(\%)} & \textbf{Diagnosis} \\
    \hline
    
    57106676 & \cite{alex_weird_2019} &  \textbf{Algo:} DQN; \textbf{Env:} CartPole-v0; \textbf{Fault:} Low update frequency of the target network. & 135.17 & 3.34 & \checkmark \\
    56964657 & \cite{alex_cartpole-v0_2019} &  \textbf{Algo:} DQN; \textbf{Env:} CartPole-v0; \textbf{Fault:} Low update frequency of the target network. & 2666.23 & 14.16 & \checkmark \\
    47750291 & \cite{kalra_deep_2017} &  \textbf{Algo:} DQN; \textbf{Env:} CartPole-v0; \textbf{Fault:} Missing exploration. & 711.21 & 19.79 & \checkmark \\
    54385568 & \cite{peterson_tensorflow_2019} &  \textbf{Algo:} DQN; \textbf{Env:} CartPole-v0; \textbf{Fault:} Wrong update of the exploration factor. & 2.87 & 17.50 & \checkmark \\
    74214034 & \cite{desert_ranger_why_2022} &  \textbf{Algo:} REINFORCE; \textbf{Env:} CartPole-v0; \textbf{Fault:} wrong computation & 169.58 & 18.17 & X \\
     & &  of the discounted reward. &  &  & \\
    45886398 & \cite{mrmjauh_function_2017} &  \textbf{Algo:} DQN; \textbf{Env:} MountainCar-v0; \textbf{Fault:} Suboptimal exploration. & 34981.4 & 11.37 & \checkmark \\
    47643678 & \cite{sheikh_dqn_2017} &  \textbf{Algo:} DQN; \textbf{Env:} CartPole-v0; \textbf{Fault:} Missing terminal state. & 3215.47 & 11.98 & \checkmark \\
    64690471 & \cite{l_pytorch_2020} &  \textbf{Algo:} DQN/DDQN; \textbf{Env:} CartPole-v0; \textbf{Fault:} not adding the end of the episode & 27.20 & 22.74 & \checkmark \\
    & &  to the replay buffer. &  &  &  \\
    67789148 & \cite{virus_dqn_2021} &  \textbf{Algo:} DQN; \textbf{Env:} CartPole-v1; \textbf{Fault:} Wrong discount factor (=1) & 59.83 & 11.97 & \checkmark \\
    56816743 & \cite{toenails_sauce_keras_2019} &  \textbf{Algo:} DQN; \textbf{Env:} CartPole-v1; \textbf{Fault:} noisy learning due to missing & 11.34 & 22.33 & \checkmark \\
    & &  the replay buffer &  &  &  \\
    49035549 & \cite{barazza_how_2018} &  \textbf{Algo:} DQN; \textbf{Env:} Pong-v0; \textbf{Fault 1:} exploration factor decays very quickly & 66.73  & 24.68 & \checkmark \\
     & &  \textbf{Fault 2:} Missing ReLU Activation &  &  & X \\

    \hline
     & &  \multicolumn{1}{r}{\textbf{Average}} & 3822.45 & 16.18 & \textbf{10/12} \\
    \hline
  \end{tabular}%
  \vspace{-1em}
\end{table*}

Table \ref{tab:table1} reports the results of debugging SO posts using \name{}. The first two columns display the post ID and its associated reference. The columns Description, Runtime, Time Overhead, and Fault Diagnosis show the fault's description, the runtime of the code sample without integrating the debugger, the time overhead added by the debugger, and whether \name{} correctly diagnosed the fault, respectively. In the Fault Diagnosis columns, \checkmark indicates that \name{} successfully diagnoses the fault, whereas \textbf{X} indicates that \name{} fails to diagnose it. 

We compared our approach's diagnosis results to the posts' accepted answers. The results demonstrate that \name{} was able to correctly diagnose 10 out of 12 faults (accuracy of 83\%). This high accuracy indicates \name{}'s effectiveness in diagnosing diverse types of faults (9 unique faults) across different DRL algorithms and environments (2 algorithms, 3 environments). In the remaining 2 faults, \name{} failed to find and diagnose the faulty behavior of the tested code. For example, in the sample \cite{desert_ranger_why_2022}, the wrong computation of the discounted reward was not detected by \name{}. Our approach does not explicitly check the discounted reward calculation since it might be computed using many valid formulas \cite{glimcher2011understanding}. Nevertheless, \name{} warned that the uncertainty was not decreasing and the reward was too distant from the maximum reward in the final episodes, indicating a problem in the learning process. In the sample \cite{barazza_how_2018}, the developer misses adding a ReLU activation function to the second and third Convolution layer of the model. \name{} does not cover these types of checks as they have already been treated by other static code analysis tools such as \cite{nikanjam2021automatic}.

Furthermore, \name{} occasionally identified additional faults that were not disclosed in the SO posts. These detections might be false positives, suggesting that \name{} could trigger unwarranted warnings. However, accurately assessing the false positive rate of \name{}'s diagnosis is challenging based on the SO posts alone, as they do not often describe every existing fault within the code. To address this, we performed a manual analysis of \name{}'s diagnostic outputs for each case, examining the relevance and correlation of these extra warnings to the primary fault identified. Overall, we found that \name{}'s additional diagnoses are linked and pertinent to the main issue. For instance, in the SO post \cite{kalra_deep_2017}, the main fault was identified as “Missing exploration” and \name{} also discovered “abnormal state entropy” and “high epistemic uncertainty”. These symptoms are consistent with a lack of exploration in DRL agents, indicating that these extra results are not false positives but potentially useful indicators of underlying issues.

\begin{conclusionbox}
\textbf{Finding 1:} \name{} performed accurately in diagnosing faulty behaviors in DRL samples extracted from SO (83\% accuracy).
\end{conclusionbox}

Furthermore, \name{}'s time overhead was assessed when debugging faulty samples. The efficiency of \name{} is a critical criterion for its adoption by developers. Table \ref{tab:table1} reported the time overhead of running the faulty samples with \name{} integrated. Running the faulty samples without \name{} took an average of 3822.5 seconds whereas running these samples with \name{} integrated added 16.2 \% of time overhead. Our approach records and monitors numerous variables such as weight, activations, observations, and rewards to perform dynamic analysis. This results in an overhead if the DRL agent or the environment is large, however, based on Table \ref{tab:table1}, this overhead is acceptable and often negligible when compared to the amount of time required to manually debug DRL systems. Moreover, \name{} is configurable, allowing the user to adjust the diagnosis periodicity to further reduce the time overhead.

\begin{conclusionbox}
\textbf{Finding 2:} \name{} exhibits a low overhead and can be leveraged to alleviate manual debugging of DRL applications.
\end{conclusionbox}

\subsection{Human evaluation}
We evaluate the usability and effectiveness of \name{} in assisting developers in finding faults in DRL code through a human study with $15$ DRL experts. We injected faults into a DRL program and recorded how many faults participants could detect with and without the \name{}.

\subsubsection{Setup}

We prepared a synthetic faulty example and conducted a one-hour individual session with each participant to assess their experience with \name{}. At the end of this session, each participant was given a survey to complete to assess \name{}’s (1) effectiveness, (2) usability, and (3) the relevance of its checks. 


We recruited 15 participants working on DRL systems (12 male, 3 female; ages 18-30) to participate in our study.
The recruitment process entailed engaging with DRL experts within our network. These recruited experts further extended our outreach by leveraging their respective networks and contacts.
The participants had diverse levels of expertise and practical experience in developing DRL applications. Their DRL experience ranged from less than a year to 5 years (median=2).
The study was conducted remotely via Google Meets video conferencing software. Two authors were always present to moderate the meeting. The Python notebook for the debugging task was shared with the participant and run with Google Colaboratory on the participant's local laptop.

For the debugging task, we followed the methodology of a previous study \cite{schoop2021umlaut}. We adapted a simple DRL program (denoted as program A) from the Pytorch tutorial \cite{noauthor_reinforcement_nodate}, which uses Deep Q-Network (DQN) \cite{mnih2013playing} to solve the Gym CartPole environment \cite{barto1983neuronlike}. We aimed to keep the program as simple as possible for two reasons: to facilitate the participant’s full comprehension and to speed up the training of the simple agent. 
To assess \name{}, we compared it with manual debugging, counting the number of faults each participant diagnosed (i.e., explicitly identifying the root causes). To establish our baseline for comparison, we first investigated a model-based static analysis tool \cite{nikanjam2022faults} recognized for detecting DRL system faults. However, its static nature restricts the investigation of dynamic, runtime behaviors required for thorough DRL diagnostics. Therefore, we chose manual debugging to accurately evaluate \name{}'s diagnostic capabilities.
We injected program A with five unique faults sourced from SO to retain the realistic error representation. While Program A operates without crashing, these faults impact the agent’s performance, notably the average cumulative reward. To simplify the debugging process and ensure fairness, we marked the locations of these faults, thereby reducing the debugging effort as the coding task is time-limited. The selected faults span essential stages of the DRL development process, such as Agent Architecture and Parameter Tuning, and vary in severity to give a comprehensive evaluation context. The following faults were introduced into Program A:

\begin{itemize}

    \item A1: Improper weight initialization for the main and target models (Low severity, model architecture) 

    \item A2: Improper bias initialization for the main and target models (Low severity, model architecture) 

    \item A3: Using the target model to predict actions instead of the main model (High severity, agent workflow)

    \item A4: Not updating the target model's parameters when reaching the update period (High severity, agent workflow)

    \item A5: The exploration factor (epsilon) value is decaying very fast (High severity, parameter tuning).

\end{itemize}

The selection of faults to be injected and the categorization of their respective severity were through consensus-driven meetings among the authors. Additionally, a pilot evaluation was performed to assess the feasibility of the proposed debugging task where two of the authors validate the appropriateness of the injected faults in terms of both severity and temporal relevance.

\subsubsection{Procedure}

Each participant was shown a presentation of our approach as well as an introductory example demonstrating the integration of \name{} into a DRL application. The code and structure of Program A were introduced before the debugging task. This familiarisation period was not included in the debugging time, ensuring that all participants began with a baseline understanding of the task and were not biased towards the faults. The participant was told that the program had numerous faults, and their task was to find them. The participant was also permitted to use any online resources, such as SO, or web search, but the moderators could not assist with debugging. 

The participant begins with the baseline condition: debugging the program without the use of the \name{}. They were given 15 minutes for debugging, and the number of faults found under the baseline conditions was counted. Next, the participant was allowed to use the \name{} for an additional 10 minutes. The number of newly found faults using the \name{} was counted at the end of the debugging time. The time allocated for the two steps was based on previous studies \cite{schoop2021umlaut} to guarantee a sufficient yet challenging window for fault detection.

At the end of the meeting, each participant was asked about their experience with \name{} and given a survey to complete. The survey includes questions regarding the tool's usability, effectiveness in assisting debugging, and the relevance of the checks integrated into \name{}.

\textbf{Bias Mitigation Strategies.} To reduce bias from gaining familiarity with Program A over time, we standardized the introductory session for all participants. This ensured each participant started with an equal understanding and time exposure to the program, allowing any increase in fault detection with \name{} to be attributed to the tool's effectiveness rather than the task familiarity. Furthermore, the purposeful allocation of less time spans for \name{} debugging than manual debugging was to demonstrate \name{}'s efficiency and effectiveness in a shorter period, balancing potential familiarity advantages earned during the manual debugging phase. Finally, immediate post-debugging surveys were undertaken to capture participants' first reactions and insights, reducing the likelihood of recall bias.

\subsubsection{Results}

\begin{table}[t]
  \centering
  \caption{Comparison between \name{} and manual debugging in terms of the number of diagnosed bugs and the average bug diagnosis time.}
  \label{tab:table2}
  \begin{tabular}{c| c c | c c | c c}
    \hline
     & \multicolumn{2}{c|}{\textbf{Manual}} & \multicolumn{2}{c|}{\textbf{\name{}}}  & \multicolumn{2}{c}{\textbf{Comparison}}  \\
     & \multicolumn{2}{c|}{\textbf{Inspection}} & &  & &  \\
     & mean  & std &  mean  & std & p-values & $\hat{A}_{12}$ \\
    \hline
    
    \# Diagnosed & 0.87&  0.96 &  3.13 &  1.31 & 0.002 & 0.91 \\
    Bugs & &  &  &  & &  \\
    \hline
    Average BD & 426.7&  203.1 &  216.3 &  125.8 & 0.03 & 0.18 \\
    time (sec)  & &  &  &  & & \\
    
    \hline
  \end{tabular}%
  \vspace{-1em}
\end{table}

\begin{figure}[t]
  \centering
  \includegraphics[width=1.05\columnwidth]{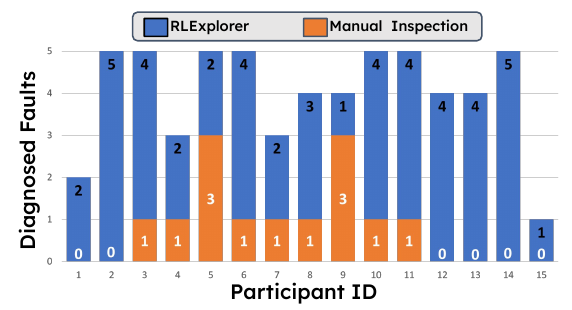}
  \caption{Number of diagnosed faults by each participant using the two debugging approaches.}
  \label{fig:fig1}
  \vspace{-1em}
\end{figure}
Table \ref{tab:table2} illustrates the consolidated results of the debugging task. Participants, when using \name{}, found and diagnose more faults ($\mu = 3.13, \sigma =1.31$) compared to the baseline condition (i.e., manual debugging) ($\mu = 0.87, \sigma =0.96$). This means that participants were able to diagnose $3.6$ times more faults using \name{} compared to manual debugging. To determine the significance and effect size of this difference, we used the Wilcoxon statistical test \cite{wilcoxon_individual_1945} and Vargha-Delaney ($\hat{A}_{12}$) \cite{vargha2000critique}. The Wilcoxon statistical test is used to examine if a difference between two means is statistically significant (which corresponds to a p-value of less than 0.05). Vargha-Delaney $\hat{A}_{12}$ is frequently used to determine the magnitude of the difference between two groups. We used the Wilcoxon statistical test, a non-parametric test, to account for the likelihood that our participants' skill levels were not normally distributed due to differences in study level and background. Results show that this difference is statistically significant (p-value = $0.002$) with a high effect size ($\hat{A}_{12}=0.91$). Furthermore, participants, when using \name{}, were able to diagnose faults in less time ($\mu = 216.3~sec, \sigma =125.8$) compared to the baseline condition (i.e., manual debugging) ($\mu = 426.7~sec, \sigma =203.1$). This difference, once again, is statistically significant (p-value = $0.03$) and has a large effect size ($\hat{A}_{12}=0.18$).

Figure \ref{fig:fig1} shows the number of faults diagnosed by each participant using the two debugging approaches. Six individuals (40\% of all participants) have failed to manually diagnose any faults. Also, no participant managed to find all faults using manual debugging alone. This demonstrates that manually debugging DRL code is very hard. On the other hand, all participants have managed to uncover at least one fault using \name{}. Nevertheless, only 47\% of all participants were able to find all 5 faults by the end of the coding exercises. This is mostly owing to the limited time given to debug the code similar to previous studies \cite{schoop2021umlaut}.
\begin{figure}[t]
  \centering
  \includegraphics[width=1\columnwidth]{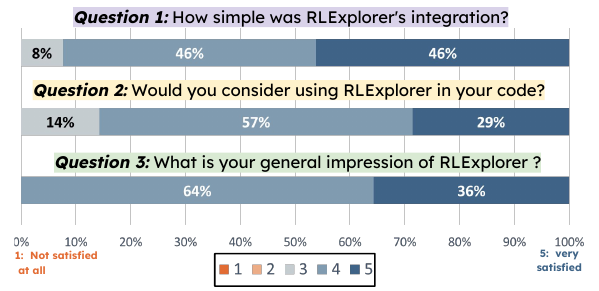}
  \caption{5-point Likert Scale Chart illustrating the feedback of participant on \name{}.}
  \label{fig:fig2}
  \vspace{-1em}
\end{figure}
Figure \ref{fig:fig2} shows the participants' replies to three 5-point Likert scale questions about the usability of \name{}. Participants were asked about their general impression of the tool, if they would consider including the \name{} in their development process, and if \name{} was easy to integrate. The replies from participants validate \name{}'s effectiveness and usefulness. Participants reported high satisfaction with the debugger and a high likelihood of leveraging \name{} in their development. Also, participants' replies confirm the  ease of integration of \name{} in DRL applications.

\begin{figure*}[t]
  \centering
  \includegraphics[width=1.8\columnwidth]{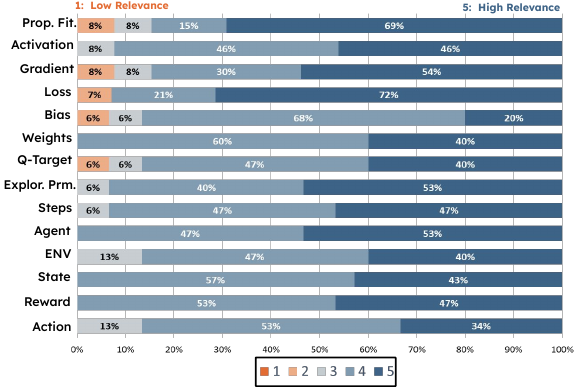}
  \caption{5-point Likert Scale Chart illustrating the feedback of participants on the relevance \name{}'s Checks.}
  \label{fig:fig3}
  \vspace{-1em}
\end{figure*}

In addition, in the survey, we described the checks contained in \name{} and asked the participants to rate the relevance and correctness of the proposed check on a 5-point Likert scale (1= low relevance, 5= high relevance). Figure \ref{fig:fig3} summarises the replies of the participants. Overall, participants found all checks relevant ($rank > 1$) with an average rating of all checks equal to 4.4 and a standard deviation equal to 0.69. On average, all checks are considered highly relevant (rank=5) by 47\% of participants, moderately relevant (rank=4) by 46\%, neutral (rank=3) by 5\%, and slightly relevant (rank=2) by 2\%. Surprisingly, the Loss check was the check with the highest average rating ($\mu = 4.57$) and was highly relevant to 72\% of participants. The common belief was that analyzing the loss in DRL is not that important compared to DL \cite{niko_loss_2017}. Yet, interviews with DRL experts and survey results demonstrate that checking the loss in DRL could be insightful. On the other hand, the Bias check was the check with the lowest average rating ($\mu = 4$) and was highly relevant to only 20\% of participants. This relatively low rating can be explained by the relatively low impact of NN-bias problems on the learning performance of DRL systems. Nevertheless, multiple participants still find this check important and hard to manually detect. In fact, the correct fine-tuning and initialization of bias values could significantly improve the learning performance of DNNs.

\begin{conclusionbox}
\textbf{Finding 3:} All participants expressed high overall satisfaction with \name{}, 92\% found it easy to use, and 86\% expressed a strong willingness to use the tool in the future.
\end{conclusionbox}

\section{Discussion}
\label{sec:discussion}
\subsection{Participants' Feedback} 

During coding interviews, we asked participants to provide open-ended feedback on \name{}'s positive and negative aspects and what might be improved. We discuss below the key participant feedback received during the human evaluation.

\textbf{\name{} checks expose silent bugs and reduce debugging time.} Many participants in the study expressed their views on the challenges associated with debugging DRL code and commented on the difficulty of identifying DRL bugs due to their silent nature, lack of explicit code failure and the high complexity of DRL systems. Therefore, participants agreed that advanced approaches like \name{} are essential to help developers save time and enhance the overall DRL development experience and highlighted how \name{} can alleviate these challenges by simplifying and expediting the diagnosis process.

\textbf{\name{} provides insightful warnings.} Several participants emphasized the relevance of \name{}’s warnings and how they can facilitate the identification and localization of faults. Participants also expressed their appreciation for \name{}'s ability to monitor and display internal signals (e.g., state entropy) during the training process in real time. One participant stated that \name{}'s plots are particularly advantageous because even in cases where warnings may produce false positives, monitoring can still provide insights into the overall learning behavior as it occurs. However, other participants requested modifications to the way the warnings were presented. These enhancements will be discussed in the next Section.

\textbf{\name{} was easily integrable  into the DRL code.} \name{}’s integration into a DRL application was meant to be easy and straightforward, requiring only a few steps and lines of code. This was validated by participants as they responded to a 5-point Likert scale question indicating high ease of integrating \name{} into the DRL application ($\mu = 4.43$). Participants' feedback was similarly encouraging, with many expressing that the simplicity of integration of \name{} would boost its chances of being embraced by the RL community.

\textbf{\name{} explicitly states that some checks might be inaccurate in some conditions.} DRL systems are black box and highly stochastic. Proposing a silver bullet check that works in all circumstances is likely impossible. We clearly explain that \name{}'s check may fail in some particular cases and may produce inaccurate warnings. The participants were aware that \name{} might overlook faults or generate false positives. They still feel that \name{} is useful. To mitigate this problem, we offered, when writing warning messages, several indications in circumstances when there is no one sure root cause. Additionally, warnings like ``high epistemic uncertainty" and ``rise in the state entropy" can still be valuable even if the training process is normal and the agent was able to realize good average reward values.

\subsection{Limitations and Future Work}

In this section, we discuss the limitations of \name{} as well as directions for future work.

\textbf{\name{} may raise false warnings (false positives) and miss errors (false negatives).} Due to the stochasticity of the DRL training, \name{} may raise false positive warnings. For example, if repeated episodes result in a low average reward value but the global average reward is stable, the ``fluctuating reward during exploitation" check might be raised. \name{} may also overlook errors (false negatives). Checks were designed to apply to general circumstances, however, these checks may not generalize to certain scenarios. Checking the Markov property of the RL environment \cite{mutti2022importance, kaelbling1996reinforcement}, for example, might be a beneficial verification for specific use cases. A deeper analysis can be used in future revisions of \name{} to improve and enhance some checks.

\textbf{Mappings from warnings to root causes may not always hold.} DRL systems are black box and highly stochastic and there are often several root causes that might result in a common error symptom. For example, a ``fluctuation in the state entropy" indicates unstable learning, however, this unstable learning might be caused by a variety of factors.

\textbf{\name{} may not generalize to new DRL algorithms.} Research is always producing new sorts of DRL algorithms and concepts, which may necessitate new debugging checks and strategies. While we rely on recent work \cite{nikanjam2022faults} to grasp the taxonomy of common DRL faults, DRL is continuously evolving, and the error trends may shift over time. We mitigate this limitation by making \name{} support the creation of custom checks that can be implemented to handle a specific problem. 

\textbf{Participants suggested improvements to \name{}’s warnings presentation.} Participants appreciated the explanations of the warning messages as they were detailed and provided actionable steps. Despite the relevance and usefulness of \name{}’s warning, participants suggested the inclusion of, summaries, HTML reports, or even web interfaces to make the debugging process more easier and appealing.

\section{Related Work}
\label{sec:related_work}
Several approaches have recently been proposed to assist developers in debugging DL systems. These approaches can be divided into two categories. The first category debugs DNN models to detect and fix faulty behaviors at the neural level (e.g., Weights and biases of the model). For example, Ma et al. \cite{ma2018mode} proposed MODE, a model debugging approach that uses state differential analysis to identify the model's internal features causing the fault and then correct it via training input selection. Similarly, Eniser et al. \cite{eniser2019deepfault} proposed DeepFault, a fault localization approach that detects the suspicious neurons in DNNs and then fixes them using gradient ascent. While these studies focus on DNN models, they can not pinpoint the source of the fault in DNN programs. The second category, closer to our approach, debugs the whole DNN program to detect and fix faulty behaviors at the code level. NeuraLint \cite{nikanjam2021automatic} and DEBAR \cite{zhang2020detecting} are two static analysis approach that detects faults in DNNs. Zhang et al. \cite{zhang2021autotrainer} proposed AutoTrainer to identify and fix DNN training faults at runtime. AutoTrainer detects five training faults: vanishing/exploding gradient, dying ReLU, oscillating loss, and slow convergence. Next, Wardat et al. \cite{wardat2021deeplocalize} proposed DeepLocalize, a DNN fault localization approach that detects root causes by spotting numerical errors during training. In the same vein, UMLUAT \cite{schoop2021umlaut} was proposed as a user interface tool to find and fix DL faults using heuristics. Amazon SageMaker Debugger \cite{rauschmayr2021amazon} is another user interface tool to detect and alert developers about frequent training faults. Most recently, DeepDiagnosis \cite{wardat2022deepdiagnosis} and DeepFD \cite{cao2022deepfd} were proposed. The first was a fault diagnosis approach that reports symptoms and proposes fixes, while the second was a learning-based fault localization approach that presents fault localization as a learning task.

To the best of our knowledge, no study has yet proposed a fault diagnosis approach for DRL systems.


\section{Threats to Validity}
\label{sec:thre-valid}


In the first part of the evaluation, we extracted real faulty samples from SO. In certain cases, the code was slightly adapted (e.g., moving from TensorFlow to PyTorch, updating obsolete packages), which may alter its logic. To mitigate this threat, the two authors cross-validated each adapted code separately and reached an agreement. 
Additionally,  in this phase of evaluation, we were unable to perform a comprehensive analysis of false positives. Distinguishing between false positives and true positives is challenging without ground truth when unreported faults are detected. To mitigate this, we performed a manual analysis of RLExplorer’s diagnostic outputs for each case, examining the relevance of additional warnings. Future work will include a systematic analysis of false positives to refine the tool's accuracy.
Another potential threat is the small sample size of participants in the human evaluation. However, the participants were experts with good RL experience (3 years of experience on average), indicating their reliable evaluation abilities. Furthermore, our methodology matches previous studies \cite{schoop2021umlaut} that interviewed a comparable number of participants. 
The design of the human study may also introduce bias. To mitigate this, we ensured that the study design was carefully structured and cross-validated by the authors. We also provided detailed instructions and a uniform environment for all participants to ensure consistency.
Finally, while developing RLExplorer, we used parameters specified by past works \cite{braiek2022testing, jones_debugging_nodate} or thoroughly fine-tuned across multiple agents/environments. These selected values may not work in some circumstances. To mitigate this threat, we allowed users to change these parameters for their specific usage easily.

\section{Conclusion}
\label{sec:conclusion}
This study proposes \name{}, the first fault diagnosis approach for DRL-based systems. Attached to the DRL application at runtime, \name{} automatically runs verification routines based on properties of the learning dynamics to detect the occurrence of DRL-specific faults. \name{}, then, displays the results of these checks as warning messages that incorporate theoretical concepts, and recommended practices. To evaluate \name{}, we collected 11 faulty DRL samples from SO and assessed \name{} on these samples. Results show that our approach was able to diagnose faults in 83\% of cases. We further conduct a human study with 15 participants to assess \name{}'s effectiveness in helping developers diagnose faults. Results show that participants were able to diagnose 3.6 times more faults using \name{} compared to manual debugging. Participants also reported high satisfaction with the debugger and a high likelihood of leveraging \name{} in their development.




\balance
\bibliography{refs}
\bibliographystyle{ieeetr}

\end{document}